 \documentclass{agnspec}
 \usepackage{graphics}
 \usepackage{psfig}

 \newcommand{\vy}[2]{#1_{\scriptscriptstyle #2}}
 \newcommand{\Ly}{Ly$\alpha$}

 \def\gtorder{\mathrel{\raise.3ex\hbox{$>$}\mkern-14mu
              \lower0.6ex\hbox{$\sim$}}}
 \def\ltorder{\mathrel{\raise.3ex\hbox{$<$}\mkern-14mu
              \lower0.6ex\hbox{$\sim$}}}
 \def\proptwid{\mathrel{\raise.3ex\hbox{$\propto$}\mkern-14mu
              \lower0.6ex\hbox{$\sim$}}}
 \def\arcsec{\ifmmode '' \else $''$\fi}
 
 \def\arcsecpoint{\ifmmode ''\!. \else $''\!.$\fi}
 
 \def\kms{\ifmmode {\rm km\ s}^{-1} \else km s$^{-1}$\fi}
 \def\Msun{\ifmmode {\rm M}_{\odot} \else M$_{\odot}$\fi}
 \def\Lsun{\ifmmode {\rm L}_{\odot} \else L$_{\odot}$\fi}
 \def\Zsun{\ifmmode {\rm Z}_{\odot} \else Z$_{\odot}$\fi}
 
 \def\ergscm2{ergs\,s$^{-1}$\,cm$^{-2}$}
 \def\icm3{{\rm cm}^{-3}}
 \def\icm2{{\rm cm}^{-2}}
 \def\qo{\ifmmode q_{\rm o} \else $q_{\rm o}$\fi}
 \def\Ho{\ifmmode H_{\rm o} \else $H_{\rm o}$\fi}
 \def\ho{\ifmmode h_{\rm o} \else $h_{\rm o}$\fi}

 \def\vFWHM{\ifmmode v_{\mbox{\tiny FWHM}} \else
             $v_{\mbox{\tiny FWHM}}$\fi}
 \def\CCF{\ifmmode F_{\it CCF} \else $F_{\it CCF}$\fi}
 \def\ACF{\ifmmode F_{\it ACF} \else $F_{\it ACF}$\fi}
 \def\Halpha{\ifmmode {\rm H}\alpha \else H$\alpha$\fi}
 \def\Hbeta{\ifmmode {\rm H}\beta \else H$\beta$\fi}
 \def\Hgamma{\ifmmode {\rm H}\gamma \else H$\gamma$\fi}
 \def\Hdelta{\ifmmode {\rm H}\delta \else H$\delta$\fi}
 \def\Lya{\ifmmode {\rm Ly}\alpha \else Ly$\alpha$\fi}
 \def\Lyb{\ifmmode {\rm Ly}\beta \else Ly$\beta$\fi}
 \def\Lyg{\ifmmode {\rm Ly}\beta \else Ly$\gamma$\fi}
 \def\hi{H\,{\sc i}}

 \def\ciii{\ifmmode {\rm C}\,{\sc iii} \else C\,{\sc iii}\fi}
 \def\civ{\ifmmode {\rm C}\,{\sc iv} \else C\,{\sc iv}\fi}

 \def\nv{N\,{\sc v}}

 \def\o5007{[O\,{\sc iii}]\,$\lambda5007$}

 \def\ovi{O\,{\sc vi}}

 \def\o{\o}
\begin{document}
 \title{  AGN Outflows: Analysis of the Absorption Troughs}
 \author{Nahum Arav\inst{1},\inst{2} }
 \institute{Astronomy Department, UC Berkeley, Berkeley, 
 CA 94720, I:arav@astron.Berkeley.EDU
 \and  Physics Department, University of California, Davis, CA 95616}
 \maketitle
 
 \begin{abstract}

With the advent of Chandra and XMM, X-ray lines spectroscopy of AGN
outflows is off to an exciting start.  In this paper we illuminate
some of the complications involved in extracting the outflow's
physical conditions (ionization equilibrium, total column density,
chemical abundances) from such spectroscopic data.  To do so we use
the example provided by high-quality FUSE and HST observations of the
outflow seen in NGC~985.

We show how simple determinations of the column density in the UV
absorption lines often severely underestimate the real column
densities.  This is due to strong non-black saturation of the
absorption troughs, where in many cases the UV line profile mainly
reflects the velocity-dependent covering factor rather than the column
density distribution.  We then show that underestimating individual
ionic column densities by a factor of 5 can cause a two orders of
magnitude error in the inferred total column density of the
outflow. In some case this will be enough to associate the UV and
X-ray absorber with the same outflowing material.

Finally, we note that the UV spectra of NGC~985 have 10--20 times the
resolution and 2--5$\times$ the S/N per resolution element compared
with the best available Chandra spectra of similar objects. Therefore,
in the X-ray band non-black saturation and velocity-dependent covering
factor effects will only become abundantly clear using vastly more
capable future X-ray telescopes. However, by taking into consideration
the lessons learned from the UV band, we can greatly improve the
quality of physical constraints we extract from current X-ray data of
AGN outflows.

 \end{abstract}

 \section{Introduction}

 Outflows in Seyfert galaxies are evident by resonance line absorption
 troughs, which are blueshifted with respect to the systemic redshift
 of their emission counterparts.  Velocities of several hundred \kms\
 (Crenshaw et~al.\ 1999; Kriss et al. 2000) are observed in UV
 resonance lines (e.g., \civ~$\lambda\lambda$1548.20,1550.77,
 \nv~$\lambda\lambda$1242.80,1238.82,
 \ovi~$\lambda\lambda$1037.62,1031.93 and \Ly), as well as in X-ray
 resonance lines (Kaastra et~al.\ 2000; Kaspi et~al.\ 2000). Similar
 outflows (often with significantly higher velocities) are seen in
 quasars that are the luminous relatives of Seyfert galaxies (Weymann
 et al.\ 1991; Korista Voit \& Weymann 1993; Arav et al.\ 2001a).
 Reliable measurement of the absorption column densities in the
 troughs are crucial for determining the ionization equilibrium and
 abundances of the outflows, and the relationship between the UV and
 the ionized X-ray absorbers (Mathur et~al.\ 1995; Elvis 2001).

 \begin{figure}
 \centerline{\psfig{figure=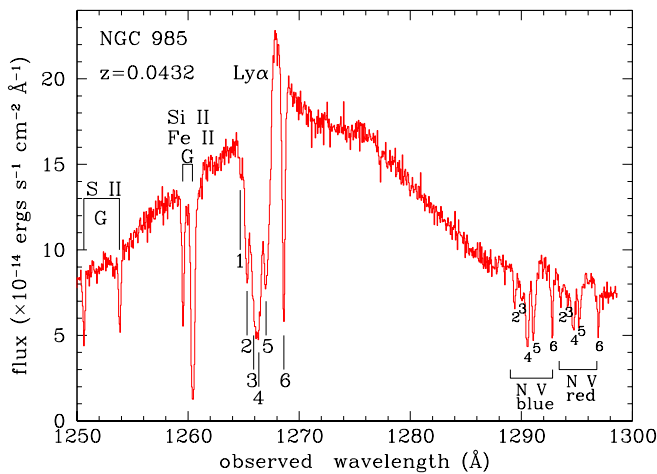,width=8.8cm,angle=-90}}
 \caption[]{HST spectrum of NGC 985, taken in 1999 February 17 with the STIS
 G140M grating (resolution $\sim10,000$), total exposure 3800 seconds.
 The absorption troughs marked with 1--6 in both \Ly\ and \nv\ 
 arise from the outflow emanating from this AGN,
   Galactic absorption is marked with a ``G.''}
 \label{985_lya_marked}
 \vspace{-0.3cm}
 \end{figure}

 In the last few years our group (Arav 1997; Arav et al. 1999a; Arav
 et~al.\ 1999b; de Kool et~al.\ 2001; Arav et~al.\ 2001) and others
 (Barlow 1997, Telfer et~al.\ 1998, Churchill et~al.\ 1999, Ganguly
 et~al.\ 1999) have shown that in quasar outflows most lines are
 saturated even when they are not black. We have also shown that in many cases
 the shapes of the troughs are almost entirely due to changes in the
 line of sight covering as a function of velocity, rather than to
 differences in optical depth (Arav et~al.\ 1999b; de~Kool et~al.\
 2001; Arav et~al.\ 2001a). As a consequence, the column densities
 inferred from the depths of the troughs are only lower limits.

 These results have led us to suspect that many of the reported column
 densities in Seyfert outflows are likely underestimated. Recently, we
 have re-analyzed the {\em HST} high resolution spectroscopic data of
 the intrinsic absorber in NGC~5548 (Arav, Korista \& de Kool 2002)
 and found that the \civ\ absorption column density in the main trough
 is at least four times larger than previously determined.
 Furthermore, we have shown that similar to the case in quasars, the
 shape of the main trough is almost entirely due to changes in
 covering fraction as a function of velocity, and is not due to
 differences in real optical depth.

 In this paper we analyze the outflow in another Seyfert~1 galaxy,
 NGC~985. Similar to the NGC~5548 case, the apparent column density (\Ly\
 in this case) is a severe underestimate of the real column density.
 Furthermore, the resulting errors in the inferred values of the total
 hydrogen column density ($\vy{N}{H}$) and the ionization
 parameter ($U$) of the absorbing gas are far larger. {\bf For the
 specific case of NGC~985, the real $\vy{N}{H}$ and $U$ values may allow
 the X-ray and UV absorbers to arise from the same gas.} We conclude
 by highlighting potential difficulties expected in the X-ray spectral
 analysis of AGN outflows, based on the UV investigation shown here and 
 elsewhere.

 \section{The case of NGC~985}

 \begin{figure}[h]
 \centerline{\psfig{figure=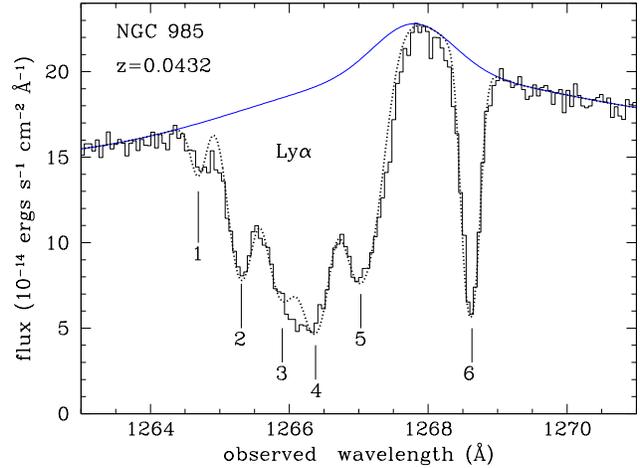,width=8.8cm,angle=-90}}
 \caption[]{\Ly\ absorption (histogram) and a simple Gaussian
 absorption model (dotted line) consisting of an emission model (solid
 line) convolved with six Gaussian absorption components.  To avoid
 excessive freedom in the model, the centroid of each Gaussian lies at
 the same velocity as the corresponding Gaussian we used in modeling
 the \nv\ absorption (see Fig~\ref{nv_abs}).}
 \label{lya_abs}
 \end{figure}

 NGC~985 is a bright ($V=13.8$ for UV see Fig~\ref{985_lya_marked},
 X-ray: $<F_{2-10kev}> = 2 \times 10^{-11}$ cgs, Nicastro et al.~1998),
 nearby Seyfert~1 galaxy, which shows a strong and complex system of
 warm absorbers ($N_H = 10^{22}$ and $10^{23}$ cm$^{-2}$) in its
 ROSAT-PSPC (Brandt et~al.\ 1995) and ASCA (Nicastro et~al.\ 1998).  In
 figure~\ref{985_lya_marked} we show the available HST data of the
 intrinsic UV absorber in NGC 985, which covers the \Ly\ and \nv\
 troughs. To find the physical conditions of the absorbing gas we first extract
 the apparent column densities for the \Ly\ and \nv\ troughs for the
 outflow, and then use these values to compute the $\vy{N}{H}$ and $U$
 of the absorbing gas.

 \subsection{Apparent column density extraction and 
 the resultant ionization equilibrium determination}

 \begin{figure}[h]
 \centerline{\psfig{figure=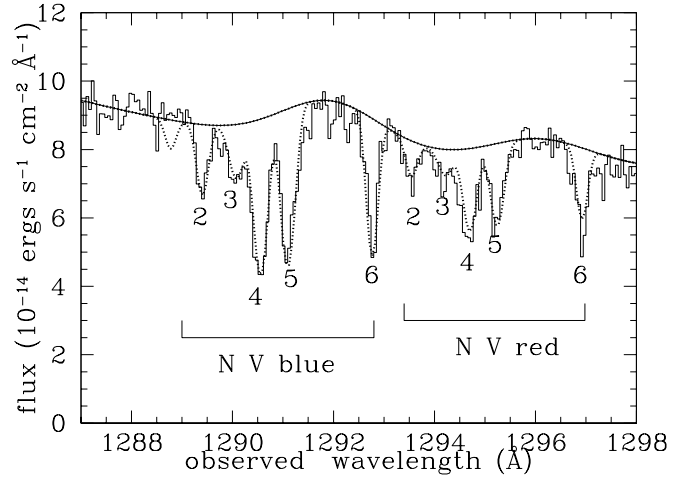,width=8.8cm,angle=-90}}
 \caption[]{A unsaturated absorption model for the \nv\ doublet.  We
 vary the Gaussian parameters to match the absorption in the blue
 doublet component and then scaled their depth by half to model the
 absorption in the red doublet component.}
 \label{nv_abs}
 \end{figure}

 Figure \ref{lya_abs} shows an unsaturated absorption model for the
 \Lya\ trough consisting of an unabsorbed emission model convolved
 with six Gaussian absorption components. The fit is very good and the
 main discrepancy (seen around 1266 \AA) is due to our restrictive
 assumption that each kinematic component for \Ly\ and \nv\ will have
 the same velocity centroid. We note in passing that this assumption
 holds quite well in this case, since four of the \Ly\ components are
 fitted very well using the velocity centroid of the \nv\
 components. For absorption component~4, the Gaussian parameters yield
 $\log(\mbox{N}_{\mbox{H~I}})=14$.

 Figure \ref{nv_abs} shows an unsaturated absorption model for the \nv\
 absorption. Components~4 and 5 are not saturated (or only slightly so)
 since their depths in the red and blue doublet lines are consistent with
 the expected 1:2 optical depth ratio (the ratio of their oscillator
 strengths). For absorption component~4, the Gaussian parameters yield
 $\log(\mbox{N}_{N~V})=14.2$

 It is usually assumed that the plasma
 is in photoionization equilibrium and therefore the relevant physical
 parameters are the total hydrogen column density ($\vy{N}{H}$) and the
 ionization parameter ($U$) of the absorbing gas.  In order to
 determine these we use the photoionization code CLOUDY (Ferland 1996)
 to try to find combinations of these parameters which will reproduce
 the measured column densities.

 \vspace{-1cm}
 \begin{figure}[h]
 \centerline{\psfig{figure=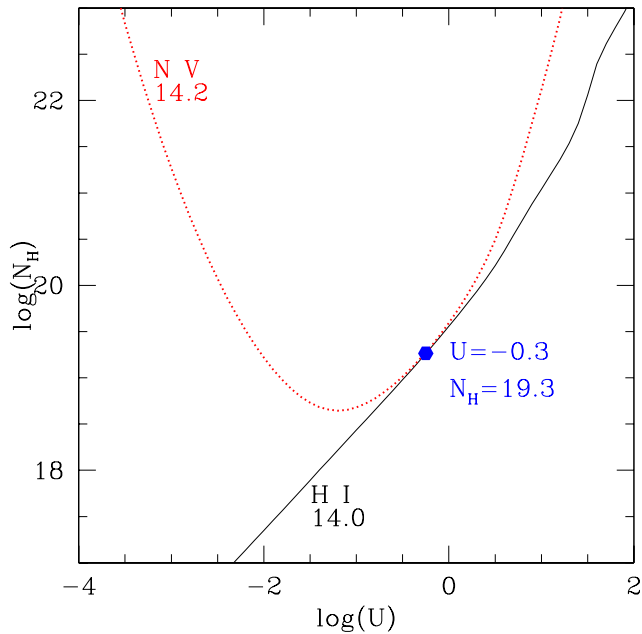,width=8.8cm,angle=0}}
 \caption[]{Ionization parameter $U$ and total column density $N_H$
 for component~4 in the NGC 985 outflow, based on the {\bf apparent}
 column densities extracted from the fits shown in figures \ref{lya_abs}
 and \ref{nv_abs}. All quantities are shown in $\log$ form.}
 \label{nh_u_hst}
 \end{figure}

 In figure \ref{nh_u_hst} we show photoionization solutions for
 absorption component~4. The presentation method is based on the grid
 models approach developed in Arav et~al.\ (2001b), using the same 
 ionizing continuum. We plot curves of
 constant $N_{ion}$ (determined from the observed troughs) in the
 $\vy{N}{H}/U$ plane. That is, within the range of plotted parameters,
 any combination of $\vy{N}{H}$ and $U$ that falls on the curve yields
 the desired column density value (for an assumed set of elemental
 abundances, solar in our case). The parameters of the absorber are
 determined by the crossing of the different $N_{ion}$ curves. We note
 that with two given curves we can obtain two solutions, since two
 crossing points are possible. By coincidence, we obtained a single
 solution in this particular case. The inferred $\vy{N}{H}$
 is at least two orders of magnitude too small to account for a typical
 warm absorber column density (Kaastra et~al.\ 2000; Kaspi et~al.\ 2000).

 \subsection{Non-black saturation}

 \begin{figure}[h]
 \centerline{\psfig{figure=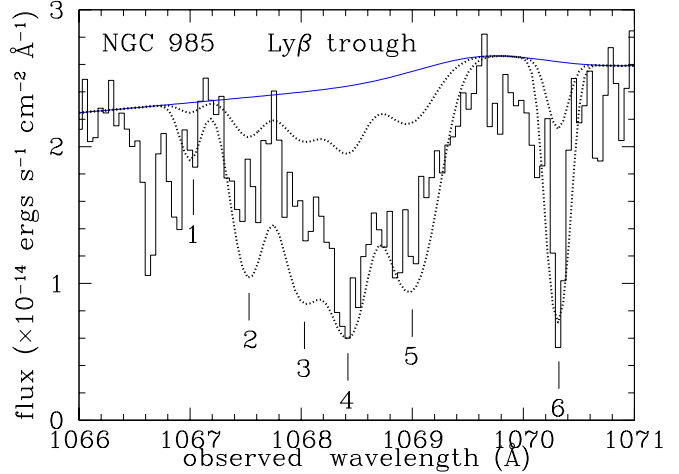,width=8.8cm,angle=-90}}
 \caption[]{Modeling the \Lyb\ absorption with the same six
 Gaussians absorption-model used in modeling the \Lya\ trough (see
 Fig.\ \ref{lya_abs}). The upper dotted line is based on assuming no
 saturation in \Lya. That is, the maximum optical depth of each component
 was multiplied by the ratios of oscillator strength and wavelength for
 \Lyb\ and \Lya, that is by $[f\lambda]_{\Lyb}/[f\lambda]_{\Lya}$. The
 poor fit in all components is indicative of saturation in \Lya. The
 lower dotted line is an absorption model assuming full \Lya\ saturation
 (i.e., using the same Gaussians with no scaling); moderate saturation
 is evident in components~4, 5 and 6 and small saturation is inferred
 for components~2 and 3.
 }
 \label{lyb_abs}
 \end{figure}

 However, our ionization equilibrium inferences are crucially dependent
 upon the reliability of the \nv\ and \Lya\ column densities inferred
 from our absorption models (Figs.\ \ref{lya_abs} and \ref{nv_abs}). The
 evidence we have collected over the last few years suggests that {\bf
 it is impossible in principle to extract a reliable column density from
 one singlet line} (see \S~1 and Arav et~al.\ 2002). It is therefore
 quite possible that our \Lya\ inferred \hi\ column density is greatly
 underestimated. Luckily we possess FUSE spectra that covers the \Lyb\
 and \Lyg\ outflow troughs in NGC~985, and can directly check the
 reliability of the \Lya\ model.

 Figure \ref{lyb_abs} shows the results of applying the \Lya\
 absorption model to the \Lyb\ trough. Our very good model-fit for the
 \Lya\ trough (Fig.\ \ref{lya_abs}) fails completely when we apply
 it to the \Lyb\ trough assuming no saturation (upper dotted line).
 A model assuming complete saturation (lower dotted line) fits the data
 considerably better, but since it over-predicts the depth of the troughs
 we determine that the \Lyb\ is not heavily saturated. This conclusion
 is confirmed by the shape of the \Lyg\ trough where all the components
 except number~6 are significantly shallower. The apparent \hi\ column
 density of component~4, as measured from the \Ly\ trough, is 5$\times$
 smaller than its real \hi\ column density deduced with the constraints
 imposed by the \Lyb\ and \Lyg\ troughs. In passing, we mention that the
 FUSE spectra also show saturation in the \ovi\ lines.

 \section{Realistic ionization equilibrium}
 \begin{figure}
 \centerline{\psfig{figure=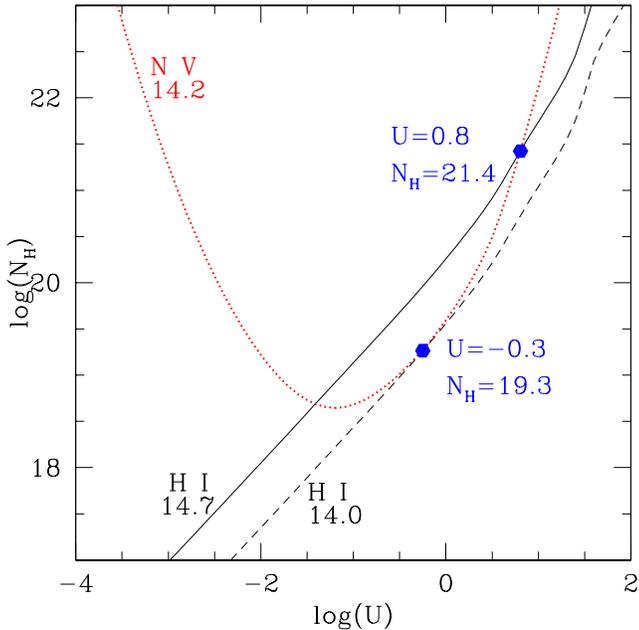,width=8.8cm,angle=0}}
 \caption[]{Similar to Fig.\ \ref{nh_u_hst} showing the ionization
 equilibrium solutions for both the real (solid line) and apparent
 (dashed line) \hi\ column density. The apparent \hi\ column density
 was extracted from the HST/STIS data alone assuming no saturation,
 while the real \hi\ column density that is five$\times$ larger than the
 apparent one is based on the FUSE data. {\bf A factor 5 difference in
 the \hi\ column density increases the inferred total absorber column
 density by more than two order of magnitudes.}  We note that the
 lower crossing-point solution (the one without the dot) can be
 excluded, since we do not detect absorption from low ionization species
 that would be present in that case.}
 \label{nh_u_hst_fuse}
 \end{figure}

 Figure \ref{nh_u_hst_fuse} illustrates the consequence of ignoring
 the possibility of saturation in \Ly\, and therefore underestimating
 the \hi\ column density in component~4 by a factor of five. The
 physical solution that uses the real \hi\ column density results in
 an absorber with 100$\times$ higher $N_H$ and roughly a 10$\times$
 higher ionization parameter than the one using the apparent column
 density. The relative difference between the two solutions is a
 robust key point of this investigation.  Absolute values of the
 correct $N_H$ and $U$ are far less accurate due to the systematic
 uncertainties in chemical abundances and incident ionizing spectrum,
 as well as the reliance on only to ionic column densities.
 Nonetheless, we point out that he solution based on realistic \hi\
 column density is compatible with physical parameters inferred for
 warm absorbers in general (Kaastra et~al.\ 2000; Kaspi et~al.\ 2000),
 while the apparent one is much too low in both $N_{ion}$ and $U$.

 \section{Summary and Relevance to analyses of X-ray warm absorbers data}

 \begin{enumerate}

 \item 

 Non-black saturation is prevalent in UV troughs of AGN outflows, even
 when the trough is fully resolved spectroscopically. This suggests that
 the same phenomenon occur in the X-ray troughs of the same
 objects. In addition, a single covering fraction per absorption
 component is a poor approximation, since covering fraction
 is often strongly velocity-dependent.

 \item Errors in column density estimates can be greatly amplified in
 the inferred ionization equilibrium for the absorbing gas. In the case
 of component~4 in NGC~985, correcting the $N_{\hi}$ by a factor of
 five resulted in a two orders of magnitude higher $N_H$ and an order
 of magnitude higher $U$. These differences are large enough to make 
 UV component~4 compatible with being the same gas as the X-ray warm absorber.

 \item Many line series are covered by the X-ray band, thus allowing
 for excellent saturation diagnostics and thus the extraction of
 reliable column densities. However, care must be taken to allow for 
 non-black saturation, which is very different from saturation
 due too  spectroscopically unresolved absorption components.

 \end{enumerate}

 \begin{acknowledgements}

 Support for this work was provided by NASA through grant number
 HST-AR-09079 from the STScI, and by NASA LTSA grant 2001-029.
 The author expresses gratitude for the hospitality of the Astronomy
 department at  UC Berkeley for the duration
 of this work, and thanks Kirk Korista for valuable discussions.

 \end{acknowledgements}


 \end{document}